\documentclass[twocolumn]{aastex63}

\received{June 1, 2019}
\revised{September 23, 2020}
\accepted{October 26, 2020}
\submitjournal{ApJL}

\shorttitle{Raman-scattered \ion{He}{2} in NGC~6886 and NGC~6881}
\shortauthors{Choi \& Lee}

\begin{document}

\title{Discovery of Raman-scattered \ion{He}{2} $\lambda$6545 in 
the Planetary Nebulae NGC~6886 and NGC~6881}

\correspondingauthor{Hee-Won Lee}
\email{hwlee@sejong.ac.kr}

\author[0000-0002-9040-672X]{Bo-Eun Choi}
\affiliation{Department of Physics and Astronomy, Sejong University, Seoul, Korea}

\author[0000-0002-1951-7953]{Hee-Won Lee}
\affiliation{Department of Physics and Astronomy, Sejong University, Seoul, Korea}



\begin{abstract}


Young planetary nebulae (PNe) retain a large amount of neutral material 
that was shed in the previous asymptotic giant branch stage. The thick \ion{H}{1}
region in young PNe can be effectively probed by illuminating far UV radiation that may
be inelastically scattered to appear in the optical region.  
Raman-scattered features are unique spectroscopic tracers of neutral regions that can be used 
to investigate the mass-loss process in young PNe. 
We conduct high resolution spectroscopy of young PNe using
BOES (the Bohyunsan Observatory Echelle Spectrograph) and report the discovery of 
a Raman-scattered \ion{He}{2} 
feature at 6545~\AA\ in NGC~6886 and NGC~6881. 
The Raman-scattered \ion{He}{2} features have been found in only five PNe so far, 
and, in particular, it is the first direct 
detection of an \ion{H}{1} component in NGC~6881. 
The Raman \ion{He}{2}~$\lambda 6545$ features in the two PNe are observed to be redshifted 
with respect to \ion{He}{2}~$\lambda6560$, indicating that the neutral regions are
expanding. We perform line profile analyses using the grid-based Monte Carlo code `STaRS'
by assuming a neutral hydrogen region in the shape of a partial spherical
shell expanding radially. 
The profiles are well fitted with the model parameters of covering factor $CF=0.3$, \ion{H}{1}
column density $N_{\rm HI} = 5 \times 10^{20}~{\rm cm^{-2}}$, and 
expansion speed $v_{\rm exp} = 25~\rm\ km~s^{-1}$ 
for NGC~6886 and  $CF=0.6$, $N_{\rm HI} = 3 \times 10^{20}~{\rm cm^{-2}}$, 
and $v_{\rm exp} = 30\rm\ km~s^{-1}$  for NGC~6881, respectively. 

\end{abstract}



\section{Introduction} \label{sec:intro}

Low and intermediate mass stars ($0.8 - 8\ M_\odot$) lose a significant fraction 
of their mass through slow stellar winds 
in the asymptotic giant branch (AGB) stage, which plays 
a crucial role in the chemical enrichment of the interstellar medium \citep{kwok05, hofner18}. 
The hot core part evolves into a white dwarf with a mass less than 
the Chandrasekhar limit ($M_{\rm WD} < 1.4\ M_\odot$), and the ejected material forms a 
planetary nebula (PN). A young PN is an ionization-bounded system which contains both ionized 
and neutral regions \citep{dinerstein91, webster88, taylor90, kastner96, huggins05, guzman-ramirez18}. 
Because young PNe have recently entered into the PN stage, they are ideal objects to 
investigate the mass-loss history in the late stage of stellar evolution.

The existence of an \ion{H}{1} component in PNe is known, but still not well understood. 
One possibility for the existence of an \ion{H}{1} component is left over from the previous AGB stage  \citep{glassgold83}. 
Another possibility is that the atomic hydrogen component is formed by the photodissociation of $\rm H_2$. 
\ion{H}{1} provides important information about the outer region of the circumstellar envelope (CSE) and the interaction 
between the CSE and interstellar medium \citep{matthews13}. 
Although atomic components are difficult to investigate due to severe confusion from 
the Galactic emission, successful \ion{H}{1} 21~cm observations have been carried out, 
resulting in successful detections of \ion{H}{1} regions with mass $\sim 0.01
\ M_\odot$ in a number of young PNe \citep{taylor90, gussie95}.

Another useful tool to probe the thick neutral region is Raman-scattered emission line features formed 
via the inelastic scattering of far-UV photons by hydrogen atoms. When a far-UV photon with 
energy near $\rm Ly\beta$ is incident on a hydrogen atom, an optical photon near H$\alpha$ 
can be emitted if the  hydrogen atom de-excites into the $2s$ state instead 
of the ground state. Raman scattering with atomic hydrogen as an astrophysical tool was first suggested 
by \cite{schmid89}. He identified the broad emission features at 6830~\AA~and 
7088~\AA~in symbiotic stars as Raman-scattered features of \ion{O}{6} $\lambda\lambda1032$ 
and 1038 \citep{allen80, akras19}. 
Symbiotic stars are wide binary systems consisting of an accreting white dwarf and a mass losing giant. They provide an ideal place to observe Raman scattering 
because of the copious amount of far UV photons emanating from the vicinity of the white dwarf. 
These UV photons are incident on a thick neutral hydrogen region formed around the evolved giant. 
\cite{nussbaumer89} proposed that photons from \ion{He}{2} can be Raman-scattered with hydrogen atoms producing emission lines that can be 
observed. These Raman-scattered \ion{He}{2} features have been detected in the symbiotic stars RR~Telescopii, V1016~Cygni, HM~Sagittae, 
and V835~Centauri \citep{vangroningen93, lee01, birriel04}. 

Strong \ion{He}{2} emitters are found amongst young PNe having a central star still 
sufficiently hot to ionize \ion{He}{2} with a thick atomic hydrogen component ejected in the previous 
AGB stage. Raman-scattered \ion{He}{2}~$\lambda6545$ was reported to be found in the 
young PNe NGC~7027, IC~5117, and NGC~6790 \citep{pequignot03,lee06,kang09}. 
The detection of Raman-scattered \ion{He}{2}~$\lambda4851$ arising from 
Raman scattering of \ion{He}{2}~$\lambda972$ was  
reported in the PNe NGC~7027, NGC~6302, NGC~6886, and IC~5117 \citep{pequignot97, groves02, pequignot03, lee06}.

In this paper, we report our first detection of Raman-scattered \ion{He}{2}~$\lambda6545$ in
the two young PNe NGC~6886 and NGC~6881. 

\section{Atomic Physics} \label{sec:atomic_phy}

Raman scattering of atomic hydrogen with incident far UV radiation near the Lyman $n\to 1\ (n\ge 3)$ series may result
in the emission of an optical photon near the Balmer $n\to2$ series. The wavelength  
of Raman-scattered radiation ($\lambda_{\rm o}$) is related to that of the incident radiation ($\lambda_{\rm i}$) 
by energy conservation:
\begin{equation}
    \lambda_{\rm o}^{-1}=\lambda_{\rm i}^{-1}-\lambda_{\rm Ly\alpha}^{-1},
    \label{eq:raman_wv}
\end{equation}
where $\lambda_{\rm Ly\alpha}$ is the wavelength of Ly$\alpha$ . 
\ion{He}{2} and \ion{H}{1} are single electron systems, sharing the same
electronic level structure. The level spacing of \ion{He}{2} is larger 
than that of \ion{H}{1} by a factor slightly exceeding 4 augmented by the fact that the
two body reduced mass of \ion{He}{2} is larger by a factor $\sim 3m_e/4m_p$, 
where $m_e$ and $m_p$ are the electron and proton masses. This leads to the far UV \ion{He}{2} ($2n\to2$) line being systematically blueshifted
by an amount $\sim -120 {\rm\ km\ s^{-1}}$ compared to the \ion{H}{1} ($n\to1$) Lyman line \citep{lee01}. 
In particular, \ion{He}{2}$~\lambda1025$ ($n=4\to2$) is Raman-scattered to form an optical feature
at 6545 \AA\ according to Eq.~(\ref{eq:raman_wv}). In addition, 
Raman scattering \ion{He}{2}~$\lambda$972 ($n=6\to2$) may form an optical feature 
at 4851 \AA.

Differentiation of Eq.~(\ref{eq:raman_wv}) yields the relation between the line widths ($\Delta\lambda_{\rm o}$ and $\Delta\lambda_{\rm i}$): 
\begin{equation}
    {\Delta\lambda_o\over\lambda_o}=
    \left({\lambda_o\over\lambda_i}\right){\Delta\lambda_i\over\lambda_i},
    \label{eq:raman_br}
\end{equation}
giving rise to the formation of broad Raman-scattered features by a factor $\lambda_{\rm o}/\lambda_{\rm i}$.
In the case of Raman scattering near Ly$\beta$, this factor is $\sim 6.4$. This means that a far UV line
with a line width $\Delta v=30{\rm\ km\ s^{-1}}$ may form a Raman feature with a significantly 
broadened width of $\sim 190{\rm\ km\ s^{-1}}$.

Cross sections for Rayleigh and Raman scattering are computed using time dependent
second order perturbation theory \citep[e.g.,][]{nussbaumer89, lee06}. In particular, 
the cross sections for Rayleigh and Raman scattering of \ion{He}{2}~$\lambda$1025 are 
$\sim 5.7\times 10^{-21}{\rm\ cm^2}$ and $\sim 7.4 \times 10^{-22}{\rm\ cm^2}$, respectively.
Raman \ion{He}{2} at 6545 \AA\ becomes an excellent probe of \ion{H}{1} regions characterized 
by the \ion{H}{1} column density  $N_{\rm HI}$ 
in excess of $10^{20}{\rm\ cm^{-2}}$.

\section{Spectroscopy of Planetary Nebulae} \label{sec:observ}
\subsection{Observation} \label{subsec:observ}

\begin{figure*}
    \gridline{
    \fig{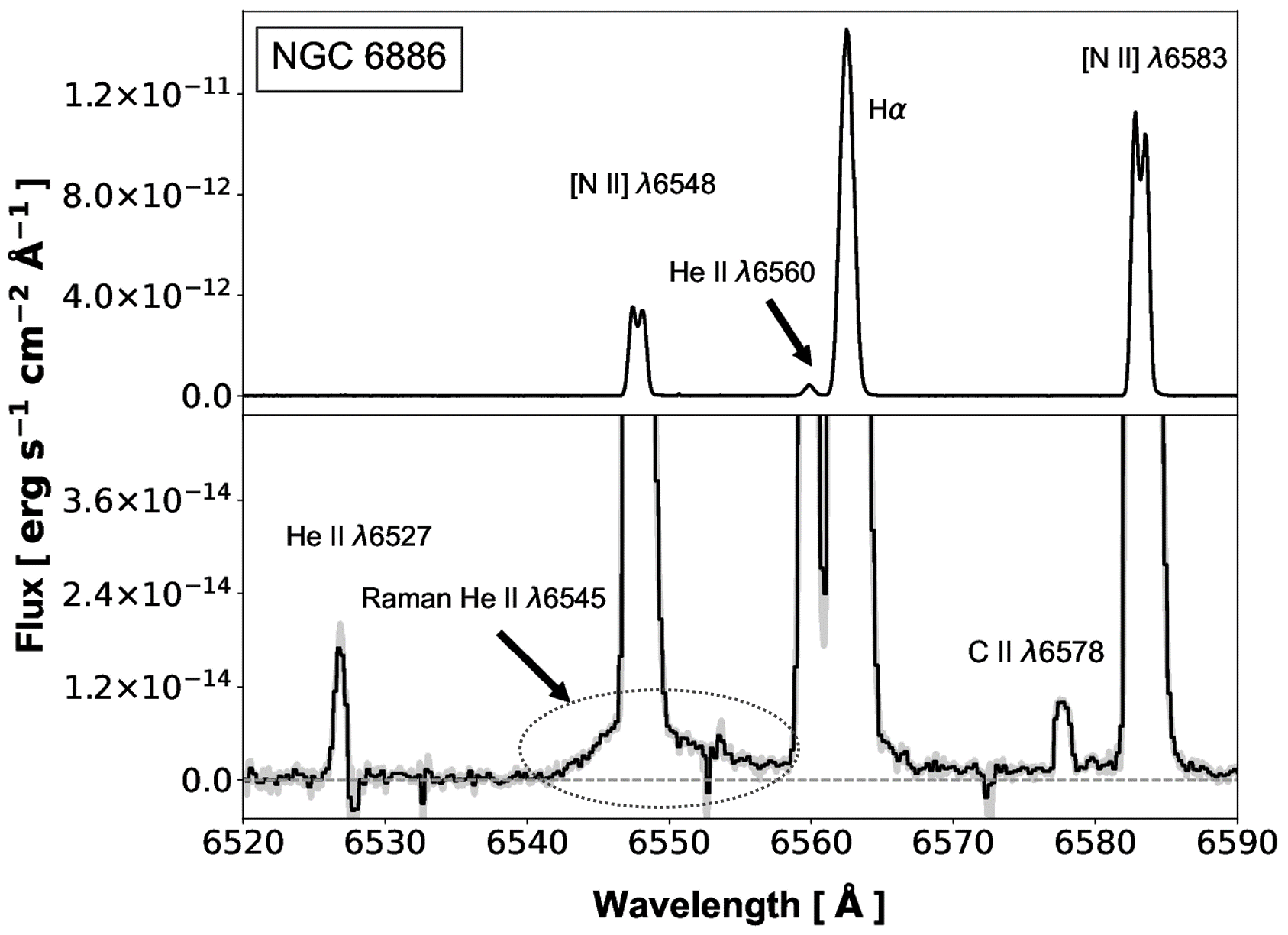}{0.5\textwidth}{(a)}
    \fig{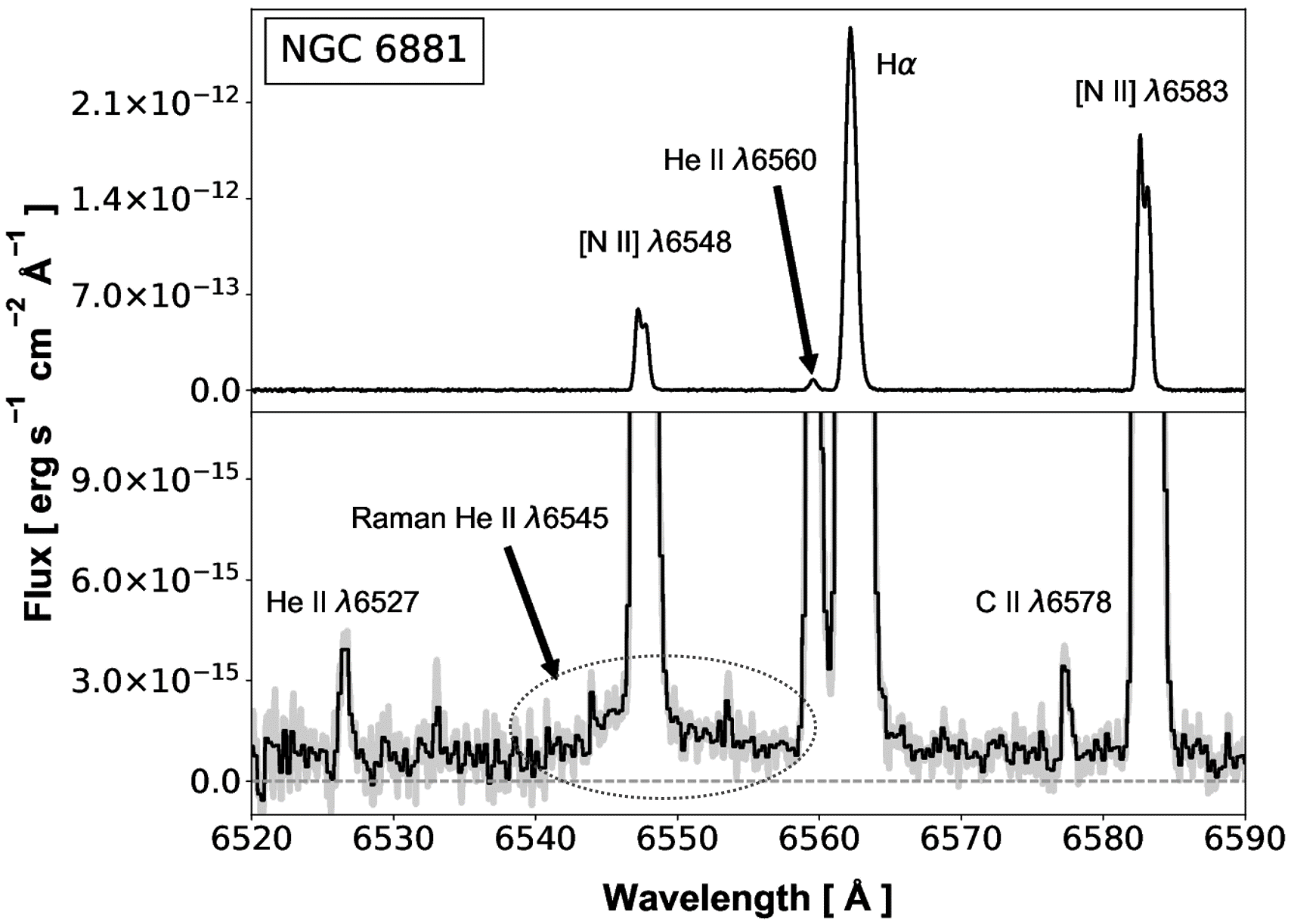}{0.5\textwidth}{(b)}
    }
    \centering
    \caption{High-resolution spectra of NGC~6886 (left) and NGC~6881 (right) obtained using BOES. 
    We present raw data (gray) and rebinned data (black). 
    Upper panels show the short exposure spectra to present strong emission lines including H$\alpha$, 
    \ion{He}{2}, and [\ion{N}{2}] lines. 
    Lower panels show the long exposure spectra rescaled for clear presentation 
    of weak lines \ion{He}{2}~$\lambda$6527, \ion{C}{2}~$\lambda$6578. The broad wing-like feature near [\ion{N}{2}]~$\lambda6548$ 
    is the Raman-scattered \ion{He}{2}~$\lambda6545$ line, and has been denoted by a circle.}
    \label{fig:boes_spectra}
\end{figure*}

Based on the \ion{He}{2} $\lambda$4686 line intensity catalog of PNe 
compiled by \cite{tylenda94}, we selected candidates with the first priority
given to strong \ion{He}{2} emitters, for which
the line intensity ratio between H$\alpha$ and \ion{He}{2}~$\lambda$4686, 
$I(HeII4686)/I(H\alpha) \gtrsim 0.05$. 
We also included the young PNe investigated by \cite{sahai11}, since they  
are considered to possess much neutral matter including atomic hydrogen 
despite a relatively low value of $I(HeII4686)/I(H\alpha)$.

We performed high resolution spectroscopy for 12 PNe during April 2019 - March 2020 
using the Bohyunsan Observatory Echelle Spectrograph (BOES)  
on the 1.8 m telescope at Bohyunsan Optical Astronomy Observatory (BOAO). 
BOES is an optical fiber-fed echelle spectrograph with a wide spectral coverage 
of 3500-10500 \AA\ encompassing the entire optical wavelength range.  
The resolution and the field of view of the spectrograph depends on the diameter of the optical fiber connected to it.
A narrower optical fiber gives higher resolution but a smaller field of view and vice versa. 
Our observations were carried out using a fiber having a 
spectral resolution of $R \sim 30,000$ and a field of view of $4.2"$. 
To improve the signal-to-noise ratio, $2 \times 2$ binning was applied at the expense of spectral resolution. 
Both short and long exposures were taken for each target because Raman-scattered \ion{He}{2}
features are very weak with a line flux about $10^{-3}$ times that of H$\alpha$.

The data were reduced using IRAF\footnote{IRAF is distributed by the National Optical Astronomy 
Observatories, which are operated by the Association of Universities for Research in Astronomy, Inc., 
under cooperative agreement with the National Science Foundation.}. Image preprocessing was carried out using bias and dome flat images. Spectra 
of a ThAr lamp were obtained for wavelength calibration. 
Flux calibration was carried out using the spectrophotometric standard stars HR~3454, HR 4554, 
HR~5501, and HR 9087.

\subsection{Spectra of NGC~6886 and NGC~6881} \label{subsec:spectra}

While there were several marginal detections of the Raman-scattered \ion{He}{2} features, here we present the two most clear detections. 
These sources were NGC~6886 and NGC~6881. 
The left and right panels of Fig.~\ref{fig:boes_spectra} show the BOES spectra of NGC~6886 
and NGC~6881, respectively. The upper and lower panels show the short and long exposure spectra, respectively. Total integrated times for 
the long exposure were 2,400 and 3,300 seconds for NGC~6886 and NGC~6881, respectively. Strong 
emission lines including H$\alpha$, \ion{He}{2}~$\lambda$6560 and [\ion{N}{2}]~$\lambda\lambda$6548
and 6583 are found in the upper panels.

In the lower panels, there appear clearly two weak emission lines \ion{He}{2}~$\lambda$6527 and \ion{C}{2}~$\lambda$6578. 
There are broad emission features seen in both sources (denoted with circles) which are blended with [\ion{N}{2}]~$\lambda$6548. 
These are the Raman-scattered \ion{He}{2} lines at 6545 \AA. 
Despite the [\ion{N}{2}]~$\lambda$6583 line being theoretically 3 times stronger than [\ion{N}{2}]~$\lambda$6548, 
the absence of a broad emission feature around the [\ion{N}{2}]~$\lambda$6583 line means we are confident in our identification of the Raman-scattered \ion{He}{2} lines. 
Furthermore, case B recombination theory predicts 
that the peak value of \ion{He}{2}~$\lambda$6527 is comparable to that of the Raman-scattered \ion{He}{2}~$\lambda$6545, 
which can be seen in Fig.~\ref{fig:boes_spectra} \citep{storey95,lee06}.

\begin{table}
\centering
\begin{tabular}{ccccc}
\hline
PN & Line & $\lambda_{\rm c}$   & $ f_0$   & $v_{\rm G}$    \\
  &       & (\AA) 
  & (${\rm erg~s^{-1}cm^{-2}\AA^{-1}}$) & ($\rm km~s^{-1}$)  \\
\hline
NGC  & H$\alpha$ & 6562.51 & $ 1.47 \times 10^{-11}$ & 50  \\
6886 & \ion{He}{2} $\lambda$6560 & 6559.84 & $4.29 \times 10^{-13}$ & 36  \\
     & \ion{He}{2} $\lambda$6527 & 6526.78 & $1.96 \times 10^{-14}$ & 36  \\
         & Raman \ion{He}{2} & 6548.65 & $7.31 \times 10^{-15}$ & 412 \\
\hline
NGC  & H$\alpha$ & 6562.23 & $3.18 \times 10^{-12}$ & 34  \\
6881 & \ion{He}{2}$\lambda$6560 & 6559.57 & $9.04 \times 10^{-14}$ & 29 \\
     & \ion{He}{2} $\lambda$6527 & 6526.63 & $4.02 \times 10^{-15}$ & 29 \\
         & Raman \ion{He}{2} & 6547.93 & $2.05 \times 10^{-15}$ & 379 \\
\hline
\end{tabular}
\caption{Single Gaussian parameters for $\rm H\alpha$, \ion{He}{2} emission lines, 
and Raman \ion{He}{2}~$\lambda6545$ feature of NGC~6886 and NGC~6881 spectra. }
\label{tab:gauss_fit}
\end{table}

The profiles of strong lines H$\alpha$ and \ion{He}{2}~$\lambda$6560 shown in the upper panels are fairly symmetric. However, it is apparent that 
the Raman \ion{He}{2} features exhibit an extended red tail. Although severe
blending with [\ion{N}{2}]~$\lambda$6548 hinders quantitative analyses, 
as a first approximation to the line profiles, we applied 
a single Gaussian fitting to each of the $\rm H\alpha$, \ion{He}{2}, and Raman \ion{He}{2}~$\lambda$6545 lines. 
The results are presented in Table~\ref{tab:gauss_fit}. 
Here, $\lambda_{\rm c}$ is observed center wavelength of each emission line, and $f_0$ is the peak value of line flux. 
The full width at half maximum (FWHM) in velocity space is represented by $v_{\rm G}$.

The line center of Raman~\ion{He}{2}~$\lambda6545$ is expected to be found at 6544.47~\AA\ for NGC~6886 
and at 6542.76~\AA\ for NGC~6881 based on the values of $\lambda_{\rm c}$ of \ion{He}{2}~$\lambda6560$
and the atomic physical relation given by Eq.~(\ref{eq:raman_wv}). However,
considerable redward deviations of $\sim +190~\rm km~s^{-1}$  
and $\sim +240~\rm km~s^{-1}$ are found for NGC~6886 and NGC~6881, respectively. 
Considering the line broadening effect of Raman scattering described by Eq.~(\ref{eq:raman_br}), 
these deviations correspond to receding motions of the \ion{H}{1} region with $\sim 30{\rm\ km\ s^{-1}}$ 
and $\sim 40~{\rm km~s^{-1}}$ with respect to the central far UV \ion{He}{2} emission region in NGC~6886
and NGC~6881, respectively.

The \ion{He}{2} emission lines are observed to be narrower than H$\alpha$ with $v_{\rm G} \sim 36~\rm km~s^{-1}$ 
for NGC~6886 and $29~{\rm km~s^{-1}}$ for NGC~6881, respectively. If the Raman conversion efficiency
is independent of wavelength, then we may expect that the Raman \ion{He}{2}~$\lambda$ 6545 
features have the line widths of $v_{\rm G} \sim 200~\rm km~s^{-1}$, which is much smaller than 
the values $\sim 400{\rm\ km\ s^{-1}}$ shown in Table~\ref{tab:gauss_fit}. However, the sharply increasing 
cross section across the \ion{He}{2}~$\lambda$1025 line implies that the Raman conversion efficiency
increases toward the red part. This leads to the resultant line profiles of Raman \ion{He}{2} that are
significantly enhanced in the red part with additional line broadening \citep{choi20}.

\section{Line Formation of Raman-scattered \ion{He}{2}} \label{sec:results}

\begin{figure}
    \centering
    \includegraphics[width=0.40\textwidth]{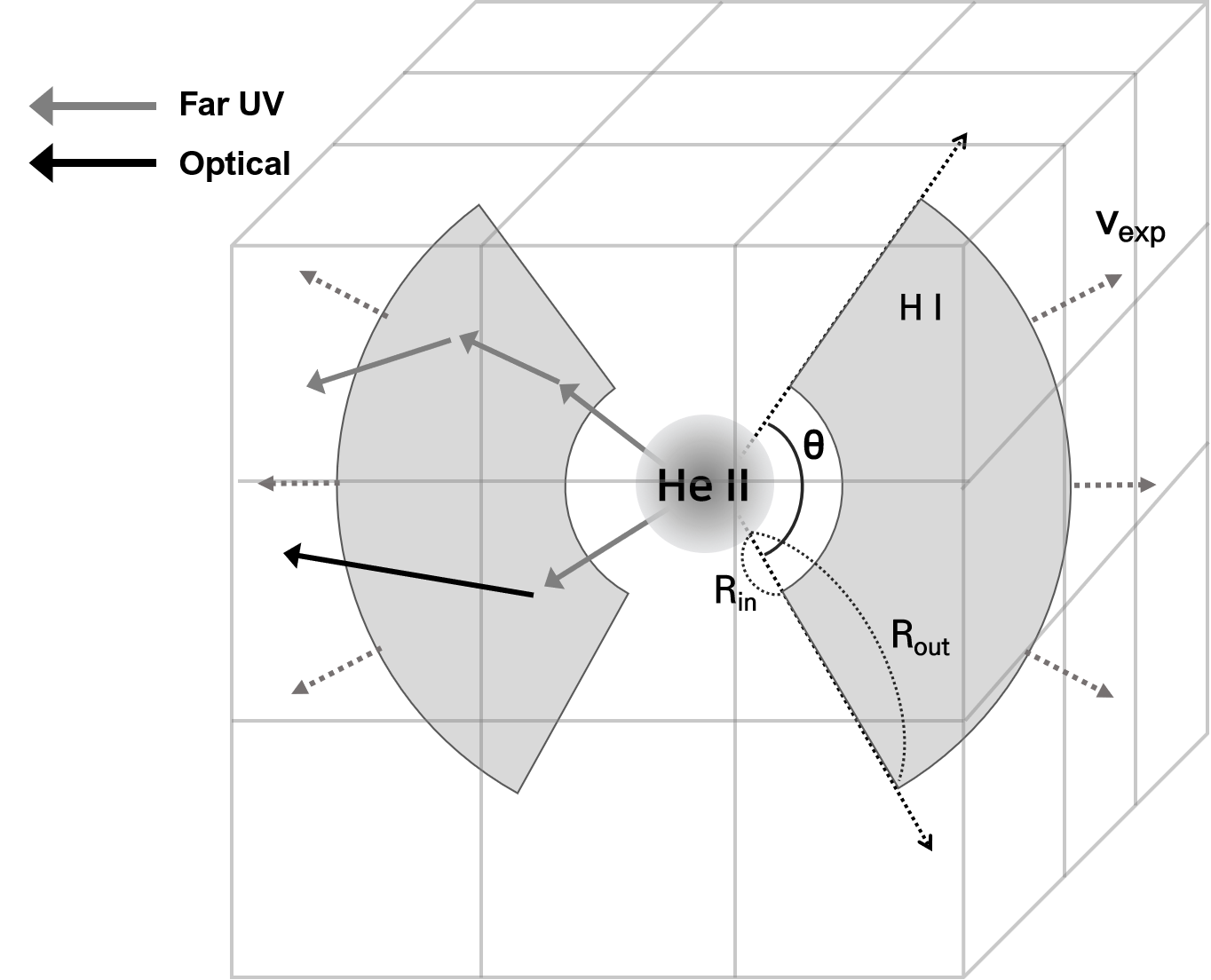}
    \caption{Schematic illustration of the simulation. The neutral region is assumed to be a partial
    spherical shell with the inner and outer radii $R_{\rm in}$ and $R_{\rm out}$, respectively, and the opening 
    angle $\theta$. The scattering region is divided into a large number of small cubical grids to trace
    each photon until escape.
    }
    \label{fig:scatt_geometry}
\end{figure}

\cite{choi20} performed an extensive study of line formation of Raman \ion{He}{2} 
in an expanding \ion{H}{1} region with a constant speed by carrying out grid-based Monte Carlo simulations. 
They showed that complicated behavior manifests itself as center shifts, significant distortions of 
the line profile, and the appearance of a secondary peak when the \ion{H}{1} region expands
with a typical speed of $\sim30{\rm\ km\ s^{-1}}$. 

The same grid-based Monte Carlo code `STaRS\footnote{http://github.com/csj607/STaRS}' is used 
to investigate the distribution and kinematics of the neutral hydrogen regions in NGC~6886 and NGC~6881. 
Fig.~\ref{fig:scatt_geometry} shows a schematic illustration of the scattering geometry adopted in this work. The
scattering region is a partial spherical shell surrounding the central \ion{He}{2} region. The covering 
factor $CF$ is defined as $CF = {\theta \over \pi}$, where $\theta$ is the opening angle of the partial 
spherical shell. The \ion{H}{1} region is assumed to be of uniform density expanding with a single 
speed $v_{\rm exp}$. 

Far UV \ion{He}{2}~$\lambda1025$ photons generated in the central \ion{He}{2} region are incident on the \ion{H}{1} region. 
The neutral region is optically thick to Rayleigh-scattered UV photons, and optically thin to Raman-scattered optical photons. 
Each far UV photon is traced until escape from the neutral region through multiple  
Rayleigh scattering, or after Raman scattering. 
The incident far UV \ion{He}{2}~$\lambda1025$ is assumed to be described by a
single Gaussian function determined in Section~\ref{subsec:spectra}.

A simulation model is set with the following free parameters: the covering factor $CF$ of the scattering region, \ion{H}{1} column density $N_{\rm HI}$, and expanding speed $v_{\rm exp}$. 
The incident \ion{He}{2}~$\lambda1025$ flux is deduced from the
observed \ion{He}{2}~$\lambda6560$ line flux and using case B recombination theory  
\citep{storey95}. 
We refer to \cite{hyung95} and \cite{pottasch05} for NGC~6886 and also \cite{kaler87} for NGC~6881  
for the values of $T_{\rm e} = 1.25 \times 10^4~\rm K$ and $n_{\rm e} = 10^4~\rm cm^{-3}$.

\begin{figure*}
    \includegraphics[width=0.95\textwidth]{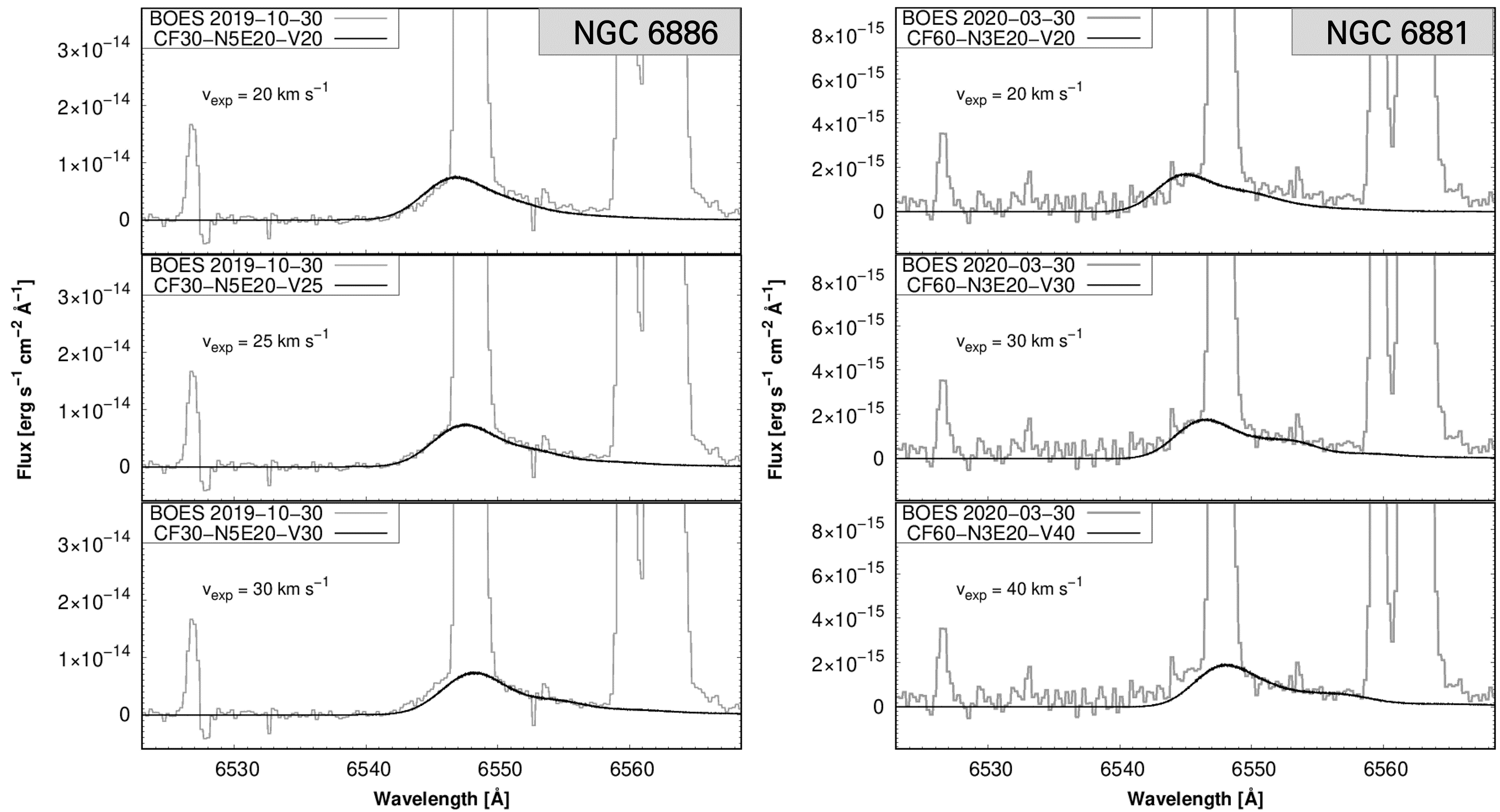}
    \centering
    \caption{
    Comparison of the observational data (gray) and our simulation data (black) for NGC~6886 (left) and NGC~6881 (right). 
    The best-fit models are found in the middle panels. 
    In the middle-left panel, the model parameters are $CF=0.3$, 
    $N_{\rm HI}=5\times10^{20}{\rm\ cm^{-2}}$, and $v_{\rm exp}=25{\rm\ km\ s^{-1}}$  
    (CF30-N5E20-V25). We also present models with a lower expansion speed of $v_{\rm exp}=20{\rm\ km\ s^{-1}}$ 
    (CF30-N5E20-V20, top-left) and a higher value of $30\rm\ km\ s^{-1}$ (CF30-N520-V30, bottom-left) 
    for comparison. In the middle-right panel, 
    the model parameters are $CF=0.6$, $N_{\rm HI}=2\times10^{20}{\rm\ cm^{-2}}$, 
    and $v_{\rm exp}=30{\rm\ km\ s^{-1}}$ (CF60-N3E20-V30). 
    The top-right and bottom-right panels show the models with 
    $v_{\rm exp}=20{\rm\ km\ s^{-1}}$ (CF60-N3E20-V20) and 
    $40{\rm\ km\ s^{-1}}$ (CF60-N3E20-V40), respectively. 
    }
    \label{fig:bestfit}
\end{figure*}

In Fig.~\ref{fig:bestfit}, we present our best-fit results for NGC~6886 and NGC~6881 in the middle panels. The gray lines represent the observational data 
and the black lines show our best-fit simulation results. Despite severe blending of
the Raman-scattered \ion{He}{2} features with strong 
[\ion{N}{2}]~$\lambda6548$ lines, the broad features with a red tail structure are conspicuous in both PNe. 
These red-tailed features are attributed to the expansion of \ion{H}{1} region.

In the case of NGC~6886, the best-fit model parameters are $CF=0.3$, 
$N_{\rm HI}=5\times10^{20}{\rm\ cm^{-2}}$, and $v_{\rm exp}=25{\rm\ km\ s^{-1}}$ 
(CF30-N5E20-V25). Similarly for NGC~6881, the best-fit model is found with parameters of $CF=0.6$, 
$N_{\rm HI}=3\times10^{20}{\rm\ cm^{-2}}$, 
and $v_{\rm exp}=30{\rm\ km\ s^{-1}}$ (CF60-N3E20-V30). 
The line flux of Raman \ion{He}{2} is mainly determined by the product of the covering
factor $CF$ and \ion{H}{1} column density $N_{\rm HI}$, resulting in degeneracy in the modeling of Raman \ion{He}{2}. However, it appears that the fit to
Raman~\ion{He}{2}~$\lambda6545$ becomes better with a somewhat lower covering factor 
$CF < 0.5$ for NGC~6886 and a higher value $CF > 0.5$ for NGC~6881, respectively. 

In Fig.~\ref{fig:bestfit}, the models with a lower expansion speed (top panels) 
and a higher speed (bottom panels) show significantly poorer fit to the observed data,
providing valid ranges of $v_{\rm exp} = 25 \pm 5~\rm km~s^{-1}$ for 
NGC~6886 and $v_{\rm exp} = 30 \pm 10~\rm km~s^{-1}$ for NGC~6881, respectively. 
However, the insufficient data quality and severe blending
with [\ion{N}{2}] prevent us from putting strong constraints on the model parameters.

\section{Summary and Discussion}\label{sec:discussion} 
We report our successful detection of Raman-scattered \ion{He}{2} at 6545 \AA\ in  
two young planetary nebulae NGC~6886 and NGC~6881. 
Gaussian model fitting is applied to find that the H$\alpha$ and \ion{He}{2}~$\lambda$6560
lines are symmetric whereas the Raman-scattered \ion{He}{2} features appear redshifted from their expected atomic line
center and exhibit an extended red tail structure indicative of
the expanding \ion{H}{1} regions with respect to the hot central emission region. 
We perform Monte Carlo simulations using the grid-based 3D code `STaRS' to obtain best fitting profiles. 

In the case of NGC~6886, \cite{pequignot03} mentioned their detection of Raman-scattered
\ion{He}{2} at 4851 \AA\ blueward of H$\beta$. Our best fit models suggest that
the column density of atomic hydrogen region $N_{\rm HI} \sim 5 \times10^{20}~\rm cm^{-2}$ 
with its covering factor $CF \sim 0.3$.  
We also measure its expanding speed $v_{\rm exp}$ to be $25~\rm km~s^{-1}$. 
A similar result was reported by \cite{taylor90}.

On the other hand, it is the first direct detection of an \ion{H}{1} component in NGC~6881. 
NGC~6881 is an interesting object exhibiting highly collimated quadrupolar lobes and a multiple ring structure 
of the ionized region that is attributed to a precessing jet \citep{guerrero98, kwok&su05}.
Furthermore, \cite{ramos-larios08} argued that its ionized quadrupolar lobes and $\rm H_{2}$ bipolar lobes show 
different morphologies and collimation degrees, indicating that this system underwent multiple bipolar ejections 
with varying mechanisms. 
We estimate the \ion{H}{1} column density $N_{\rm HI} \sim 3 \times 10^{20}~\rm cm^{-2}$ 
and the expansion speed $v_{\rm exp} = 30~\rm km~s^{-1}$ assuming a covering factor of $CF \sim 0.6$.

The total \ion{H}{1} mass $M_{\rm HI}$ based on our model is given by
\begin{equation}
    M_{\rm HI} \simeq 1.4 \times 10^{-4}~\left({N_{\rm HI} \over 10^{20}~\rm cm^{-2}}\right) 
               \left({R_{\rm out} \over 10^3~\rm au}\right)^2 CF~M_{\odot}, 
    \label{eq:hmass}
\end{equation}
where we set $R_{\rm out}$ to 2 times of $R_{\rm in}$. 
The distance to NGC~6886 is about 2.6~kpc and the angular size of its central ionized region 
as measured from its optical image is roughly $5''$ \citep{pottasch05} (assumed to be $R_{\rm in}$), corresponding to a linear size of 13,000 au. 
This leads to an estimate of $M_{\rm HI} \sim 0.03~M_{\odot}$. 
The distance to NGC~6881 is $\sim 2.5~\rm kpc$ \citep{cahn92} with an angular size of its central ionized region being 
$\sim 5''$ \citep{kwok&su05}, corresponding to a linear size of 13,000 au and therefore $M_{\rm HI} \sim 0.04~M_{\odot}$.
Setting $R_{\rm out}$ to be twice $R_{\rm in}$ is somewhat arbitrary and of course if we set $R_{\rm out}$ to be 4 times $R_{\rm in}$, then the \ion{H}{1} mass estimate 
will increase by 4 times.

Raman-scattered \ion{He}{2} features can be formed near the hydrogen Balmer series at 6545~\AA, 4851~\AA, and 4332~\AA~
derived from Raman scattering of far UV \ion{He}{2} $\lambda 1025$, $\lambda 972$, and $\lambda 949$, respectively. 
The integrated line analyses of Raman \ion{He}{2} features are an appropriate probe of \ion{H}{1} regions  
with a column density of $N_{\rm HI}\sim 10^{20-23}~\rm cm^{-2}$. Deep spectroscopy is required to detect Raman \ion{He}{2}~
$\lambda\lambda 4851$ and 4332. The detection of these lines will allow us to carry out 
better studies of the kinematics and distribution of thick neutral hydrogen regions in PNe.

It is notable that both Raman \ion{He}{2} features and \ion{C}{2}~$\lambda6578$ in emission is detected in NGC~7027, IC~5117, 
NGC~6790, similar to our two targets NGC~6886 and NGC~6881 
\citep{keyes90, lee06, kang09}. Furthermore, these objects are also strong polycyclic aromatic hydrocarbon (PAH) emitters 
\citep{smith08, ohsawa16} indicating that these PNe possess carbon enriched environments from the previous AGB stage 
as a result of dredge-up processes. However, due to small number statistics it is too early 
to conclude that Raman-scattered \ion{He}{2} features are found in only carbon enriched PNe.
A definite conclusion should wait for a more systematic survey of PNe with Raman-scattered \ion{He}{2}.

\acknowledgments
The authors are grateful to the staff of the BOAO and Seulgi Kim for her help in the data acquisition of NGC~6886. 
This research was supported by the Korea Astronomy and Space Science Institute under the R\&D program (Project No. 2018-1-860-00) 
supervised by the Ministry of Science, ICT and Future Planning. 
This work was also supported by a National Research Foundation of Korea (NRF) grant funded by 
the Korea government (MSIT: No. NRF-2018R1D1A1B07043944). 





\end{document}